\documentclass{elsart}
\usepackage[square,sort,comma,numbers]{natbib}
\usepackage[dvips]{graphicx}
\usepackage{amssymb}

\begin{document}
\begin{frontmatter}

\title{PINO - a tool for simulating neutron spectra resulting from the $^7$Li(p,n) reaction}

\author[GSI,GUF]{R.~Reifarth}, 
\ead{r.reifarth@gsi.de}
\author[GSI]{M.~Heil}, 
\author[FZK]{F.~K{\"a}ppeler} ,
\author[GSI,GUF]{R.~Plag} 
\address[GSI]{Gesellschaft f{\"u}r Schwerionenforschung mbH, Darmstadt, D-64291, Germany}
\address[GUF]{J.W. Goethe Universit{\"a}t, Frankfurt a.M, D-60438, Germany}
\address[FZK]{Forschungszentrum Karlsruhe, Postfach 3640, D-76021 Karlsruhe, Germany}

\maketitle
\begin{abstract}
The $^7$Li(p,n) reaction in combination with a 3.7~MV Van de Graaff accelerator 
was routinely used at FZK to perform activation as well as time-of-flight measurements with 
neutrons in the keV-region. 
Planned new setups with much higher proton currents like SARAF and FRANZ and the availability of liquid-lithium target 
technology will trigger a renaissance of this method. A detailed understanding of the neutron
spectrum is not only important during the planning phase of an experiment, but also during for the analysis
of activation experiments. 
Therefore, the Monte-Carlo based program PINO (Protons In Neutrons Out) was developed, 
which allows the simulation
of neutron spectra considering the geometry of the setup and the proton-energy distribution.
\end{abstract}


\end{frontmatter}

\section{Introduction}
The scope of this work was to provide a tool for experimentalists to estimate the neutron flux as a well
as the neutron energy distribution during activation experiments using the $^7$Li(p,n) reaction as a neutron source.
This method in combination with a Van de Graaff accelerator was used for almost thirty 
years at the Forschungszentrum Karlsruhe. A detailed description of the activation method for obtaining  
(n,$\gamma$) cross sections and the experimental setup can be 
found in literature \cite{BeK80,TDK95}. The essential features of such experiments 
consist of two steps, irradiation of a sample in a 
quasi-stellar neutron spectrum and the determination of the amount of freshly produced nuclei either  
via the induced activity \cite{ReK02} or via accelerator mass spectrometry (AMS)~\cite{NPA05}.
The development of new accelerator technologies, in particular the development of radiofrequency quadrupoles (RFQ) 
provides much higher proton currents than previously achievable. The additional 
development of liquid-lithium target 
technology to handle the target cooling opens a new era of activation experiments thanks to the enormously 
increased neutron flux. 
Projects like SARAF~\cite{NMB06} and FRANZ~\cite{MCM06},
which are currently under construction underline this statement. Even though FRANZ allows also the
preferable time-of-flight (TOF) method, the activation method, if applicable, remains the method of choice
if the half-life of the isotope under investigation is too short or the sample mass too small. Many of the 
astrophysically interesting isotopes will have to wait for neutron sources even beyond SARAF or FRANZ, if 
a time-of-flight measurement is desired \cite{CoR07}.

While other neutron-energy distributions were used on occasion \cite{RSK00,RHF08}, 
the quasi-stellar neutron spectrum, which can be obtained by bombarding a 
thick metallic Li target with protons of 1912~keV, slightly above 
the reaction threshold at 1881~keV, was the working horse at the Forschungszentrum Karlsruhe 
\cite{BBK00}. Under such conditions, the $^7$Li(p,n)$^7$Be reaction 
yields a continuous energy distribution with a high-energy 
cutoff at $E_n$~=~106~keV. The produced neutrons are emitted in a 
forward cone of 120$^\circ$ opening angle. The angle-integrated 
spectrum closely resembles a spectrum necessary to measure the Maxwellian averaged cross
section at $kT$~=~25~keV:

\begin{equation}
  \frac{dN}{dE} = E \cdot e^{-\frac{E}{kT}} = \sqrt{E}\cdot \Phi_{Maxwell},
\end{equation}

where $\Phi_{Maxwell}$ is the Maxwellian distribution for a thermal energy of 
$kT$~=~25~keV \cite{RaK88}.

\begin{figure}
  \includegraphics[width=.9\textwidth]{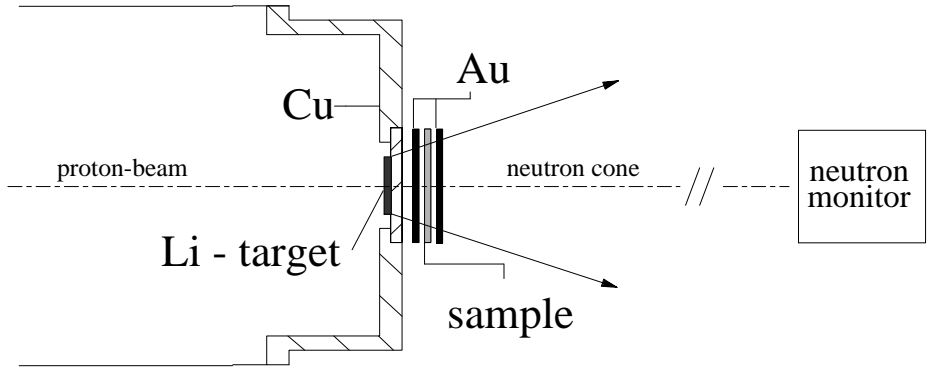}
  \caption{Typical activation setup at the Forschungszentrum Karlsruhe. Neutrons are produced via the $^7$Li(p,n) reaction
  just above the production threshold. The emitted neutrons are then kinematically focussed into a cone with an opening angle of
  120$^\circ$. The sample is usually sandwiched by two gold foils in order to determine the neutron flux just before 
  and behind 
  the sample. 
  \label{activation_setup}}
\end{figure}

The samples are typically sandwiched between gold foils and placed 
directly on the backing of the lithium target. A typical setup is 
sketched in Fig.~\ref{activation_setup}. 
The simultaneous activation of 
the gold foils provides a convenient tool for measuring the 
neutron flux, since both the stellar neutron capture cross section
of $^{197}$Au \cite{RaK88} and the parameters of the 
$^{198}$Au decay \cite{Aub83} are accurately known.

While the neutron spectrum for the standard case is very well understood, a tool 
for extrapolation to different experimental conditions is desired. Such changes of the
standard setup typically include differences in the angle coverage of the sample, a different
thickness of the lithium layer, or different proton energies. The extrapolation is, while 
conceptually obvious, not straight forward. After impinging onto the lithium layer,
the protons are slowed down until they either leave the lithium layer (in case of a very thin layer)
or are below the (p,n) reaction threshold and do not contribute to the neutron production anymore.
The double-differential (p,n) cross section changes significantly during this process, 
especially in the energy regime close to the production threshold. Additionally the kinematics of
the reaction is important during the process. Since the Q-value of the reaction is positive, the 
reaction products, and the neutrons in particular, are emitted into a cone in the direction 
of the protons (Fig.~\ref{activation_setup}). This effect becomes less and less pronounced as the 
proton energy increases. If the proton energy in the center-of-mass system is above 2.37~MeV, a
second reaction channel $^7$Li(p,n)$^7$Be$^\star$ opens, which leads to a second neutron group 
at lower energies.

To model these processes quantitatively, a tool to simulate the neutron spectrum resulting 
from the $^7$Li(p,n) reaction with a Monte-Carlo approach is indispensable. Therefore we 
developed the highly specialized program PINO - Protons In Neutrons Out.     
  
\section{PINO - the program} 

\subsection{General approach \label{approach}}

The Monte-Carlo approach means that for each neutron, at first a randomly chosen energy 
will be assigned to a proton. This happens according to the energy distribution of the proton beam.
Then the proton energy at the event of an interaction in the lithium layer will be 
randomly determined. If $E_p$ is the proton energy and $z$ the depth inside the lithium layer,
the stopping power is defined as:

\begin{equation}\label{stoppingpower}
		S=-\frac{dE_p}{dz}		
\end{equation}

The the proton energy as a function of position inside the lithium layer can then be written as:

\begin{equation}
		E_p(z)=E_p(0) - \int_0^z{S(E_p(z'))dz'}		
\end{equation}

or the depth as a function of energy using the projected range $z(E_p,E_p(0))=R(E_p(0))-R(E_p)$.
For the interaction probability as a function of depth one finds 
${dP}/{dE}(z) \propto \sigma_{(p,n)}(E_p(z))$, or as a function of energy:		

\begin{equation}
		\frac{dP}{dE}(E) \propto \frac{\sigma_{(p,n)}(E)}{S(E)}.		
\end{equation}

Hence the total interaction probability is:

\begin{equation}\label{ptot}
		P_{total}(E_p) \propto \int_{E_{threshold}}^{E_p} \frac{\sigma_{(p,n)}(E)}{S(E)} dE .		
\end{equation}

After the determination of the interaction energy, the emission angle and energy are randomly acquired 
based on the double-differential $^7$Li(p,n) cross section fulfilling
momentum and energy conservation laws. Finally the neutron will be tracked through the imaginary sample.

\subsection{Proton transport \label{p_transport}}

Protons are not fully tracked. The simplified assumption was made that the protons do not scatter in 
the thin lithium layer. The only interactions considered are the ionization of lithium, hence the stopping 
of the protons because of electronic interactions, and (p,n) reactions. 
The stopping power data (see equation~(\ref{stoppingpower})~) are read in from a separate file, which 
can be initialized at the beginning of the program (see Sect.~\ref{input_output}). 
All results shown in this paper are based on stopping power 
data from the well-established program SRIM \cite{Zie80}.

\subsection{The (p,n) reaction \label{reaction}}

Under the conditions typically simulated here, 
the total interaction probability is in the order of $10^{-6}$.
In order to simplify the calculations it is therefore assumed that the total interaction 
probability is much smaller than one. In other words, the self-shielding
of the lithium layer is neglected. 
The advantage of this assumption is, that the (p,n)  
cross section can then be scaled such that for each proton exactly one 
neutron is produced in the simulation. This
scaling factor is the total neutron production yield $P_{total}$, see equation~(\ref{ptot}). 
The interaction probability used in the code is then: 

\begin{equation}
		\frac{dP_{code}}{dE}(E) = \frac{1}{P_{total}} \frac{\sigma_{(p,n)}(E)}{S(E)},		
\end{equation}

hence $P_{code,total}(E_p) = 1$.		

The cross section data are read in from a separate file, which 
can be initialized at the beginning of the program (see Sect. \ref{input_output}). 
All results shown in this paper are based on the double-differential cross sections compiled by 
Liskien and Paulsen \cite{LiP75}.

\subsection{Neutron transport \label{n_transport}}

Similarly to the protons, also the neutron transport is very simplified. In the typical applications considered here, 
the backing of the lithium layer as 
well as the sample were thin enough to result in transmission coefficients above 95\%. 
The neutrons are therefore not tracked, 
but rather ray-traced. Only position and angle at the time of production are considered.  

\subsection{Input and Output}
\label{input_output}

Each simulation requires an input file, which contains information about the geometry as well as cross
sections and output options. The first three lines of the input file refer to the 
files containing the double-differential $^7$Li(p,n)$^7$Be and $^7$Li(p,n)$^7$Be$^\star$ cross sections
and the stopping power data. Modification of these input parameters allows for instance the simulation of
chemical compounds of lithium, like the thermally very robust compositions LiF or Li$_2$O. 

The next lines define the proton energy and energy spread, thickness of the lithium layer, and the number 
of samples. Finally, the last lines are defining the geometry for the emitted neutrons and contain information
about the size of the lithium spot hit by the proton beam, the distance from sample to lithium layer, 
and the geometry of the sample (shape, dimensions). For convenience, an optional line allows a reference to 
a neutron capture cross section file and the resulting neutron spectrum will be folded with this cross section. 

At the end of each simulation two output files are generated. One contains neutron yields as a function of 
energy. Only neutrons passing the sample are considered and output is generated for $^7$Li(p,n)$^7$Be 
and $^7$Li(p,n)$^7$Be$^\star$ separately. Additionally, both outputs are given with an angular weighting. If
neutrons pass a disk not perpendicular to the surface, the effective sample thickness changes 
by a factor cos($\vartheta$). 
The other file contains a suite of useful information for the planning and analysis of an experiment. Amongst them
are the number of neutrons passing the sample, the number of neutrons per proton current, the effective 
lithium thickness, the particular layer after which protons are below the neutron production threshold, 
and the neutron 
capture cross section
folded with the neutron spectrum derived for the sample.    

\subsection{Limitations and possibilities}
As already discussed in Secs.~\ref{p_transport} through \ref{n_transport} the particle transport is simplified. In particular
straggling of the protons in the lithium layer is not considered. This is a valid assumption
for the typical applications described in 
Sec.~\ref{results}, where the range of the protons above the neutron production threshold is very small. If however, 
one would choose significantly higher proton energies and respectively thicker lithium layers, the proton
straggling will lead to a more isotropic neutron emission than predicted by the program, since the protons might
change direction before interacting with the $^7$Li nucleus.

Similarly the simplified neutron transport is only valid for the cases described before. 
If the interaction probability in the sample or 
backing can not be neglected anymore, because of thicker backings or samples, the simulation 
may result in significant differences from reality. 

If however, the described simplifying assumptions are valid, the program is very flexible, easy to use, and
very fast. A typical simulation with 10$^9$ protons
on a standard Laptop (Intel processor, 2~GHz) takes only minutes, which 
means that situations with 10$^9$ neutrons can be simulated. 
Depending on the problem however, 10$^7$ emitted neutrons are often
already sufficient, resulting in simulation times of a few seconds.
PINO will therefore be made available as a web-application at the URL: http://exp-astro.physik.uni-frankfurt.de/pino 

\section{PINO - results \label{results}}
\subsection{Comparison with measured data}

A first test of the performance was the comparison with the experimentally determined 
neutron spectrum at the Forschungszentrum Karlsruhe by Ratynski and K{\"a}ppeler \cite{RaK88}.
We feel, it is important to point out that the spectrum measured and published there 
is not the spectrum seen by a disk-like sample. In the spectrum measurement, the same neutron detector was
positioned at different angles with respect to the proton-beam axis. The surface 
of this detector was always perpendicular to the direction of the emitted neutrons. Therefore, the 
effective thickness of the detector was constant for all neutron emission angles. In contrast,
the effective thickness of a disk-like sample changes as a function of emission angle (see also Sec.~\ref{input_output}). 

\begin{figure}
  \includegraphics[width=.9\textwidth]{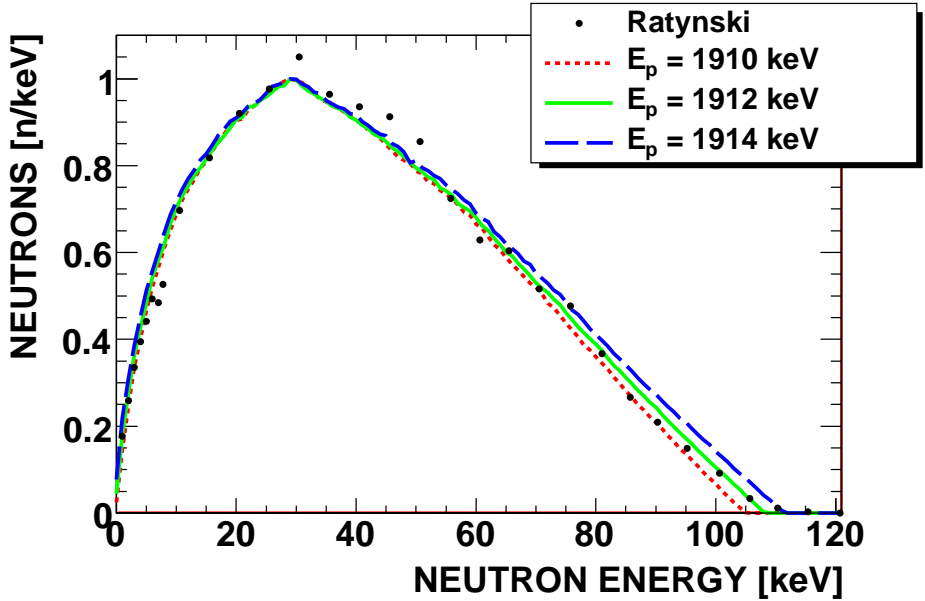}
  \caption{Comparison of the number of angle-integrated neutrons per linear energy bin for measured data \cite{RaK88} 
  with simulation results that contain no weighting with the 
  angle-dependent sample thickness. All simulated spectra are normalized to a common maximum of 1. 
  \label{comparison_ratynski_no_w}}
\end{figure}

Fig.~\ref{comparison_ratynski_no_w} shows the comparison of the experimental data with simulations of different
incident proton energies. The data agree very well for the nominal proton energy of 1912~keV as well as for energies
slightly above or below. This gives additional confidence in the experimental method, since the proton energy might fluctuate 
slightly during the sometimes extended neutron activation experiments.    

\begin{figure}
  \includegraphics[width=.45\textwidth]{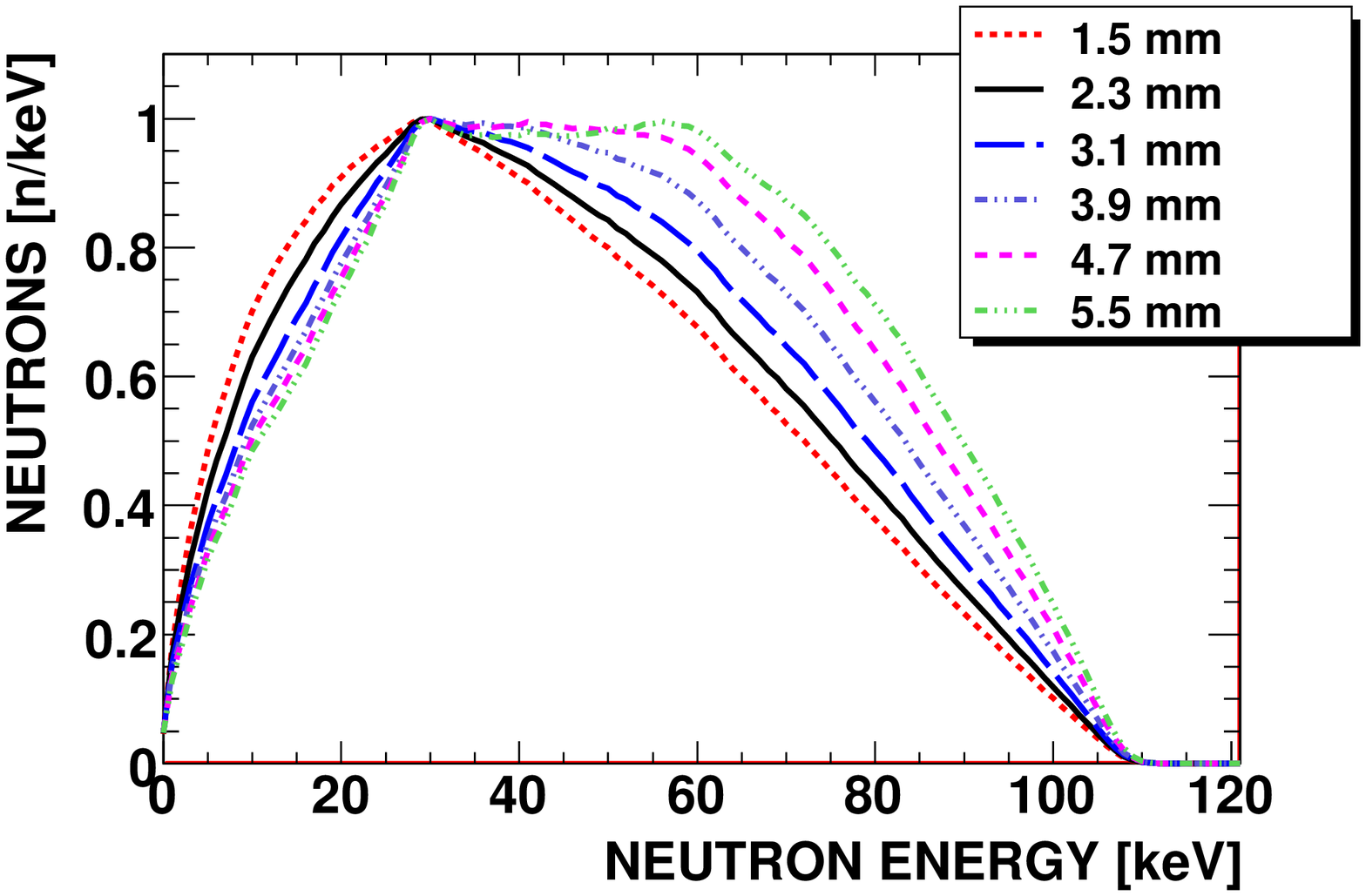}
  \includegraphics[width=.45\textwidth]{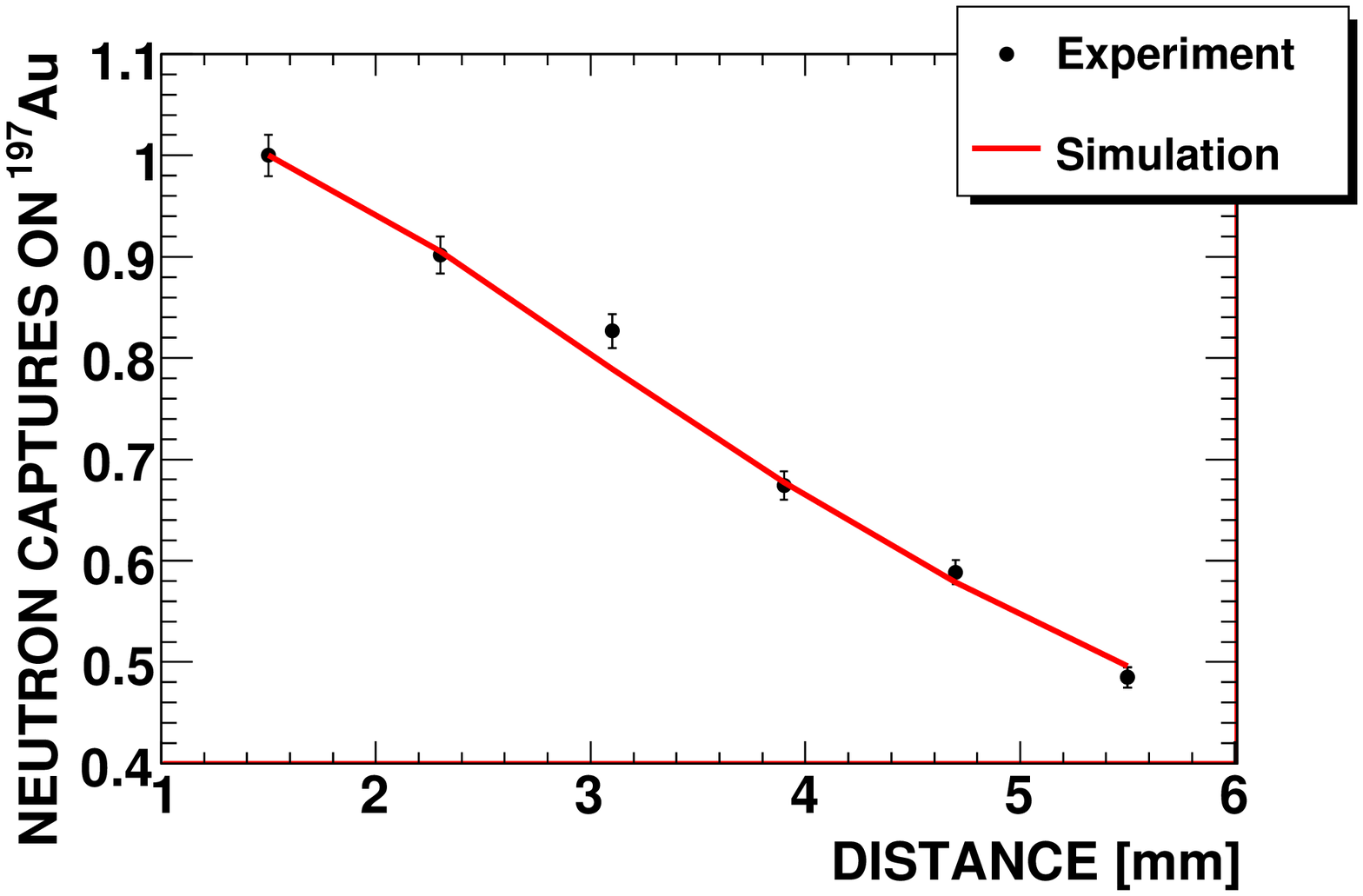}
  \caption{Left: Simulated neutron spectra for the stack of gold foils with 5~mm diameter (see also Fig.~\ref{comparison_ratynski_no_w}). Right: Comparison of
  experimental data with simulation results. The data are normalized to the point at 1.5~mm.
  \label{comparison_stack_5mm}}
\end{figure}

\begin{figure}
  \includegraphics[width=.45\textwidth]{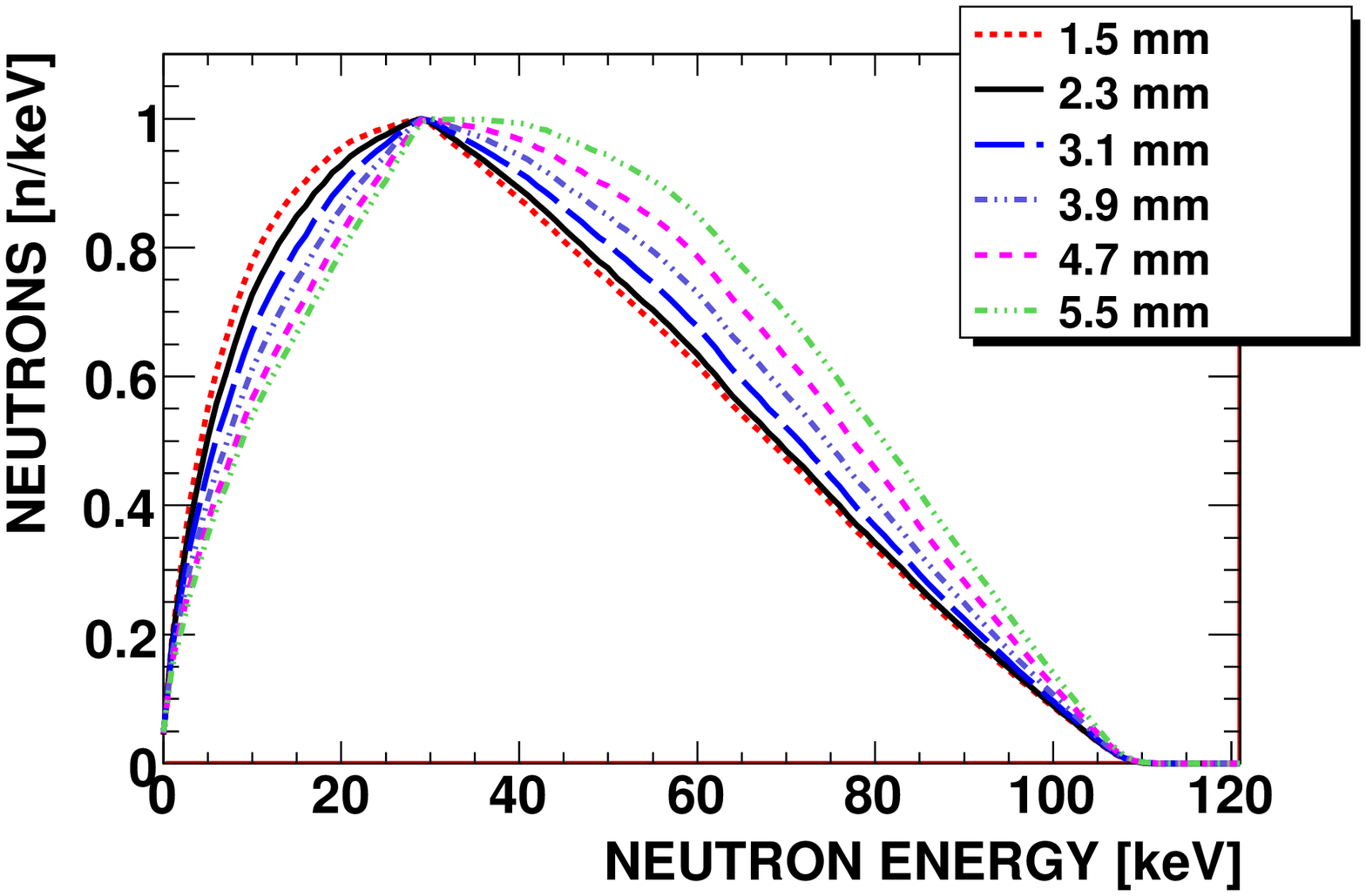}
  \includegraphics[width=.45\textwidth]{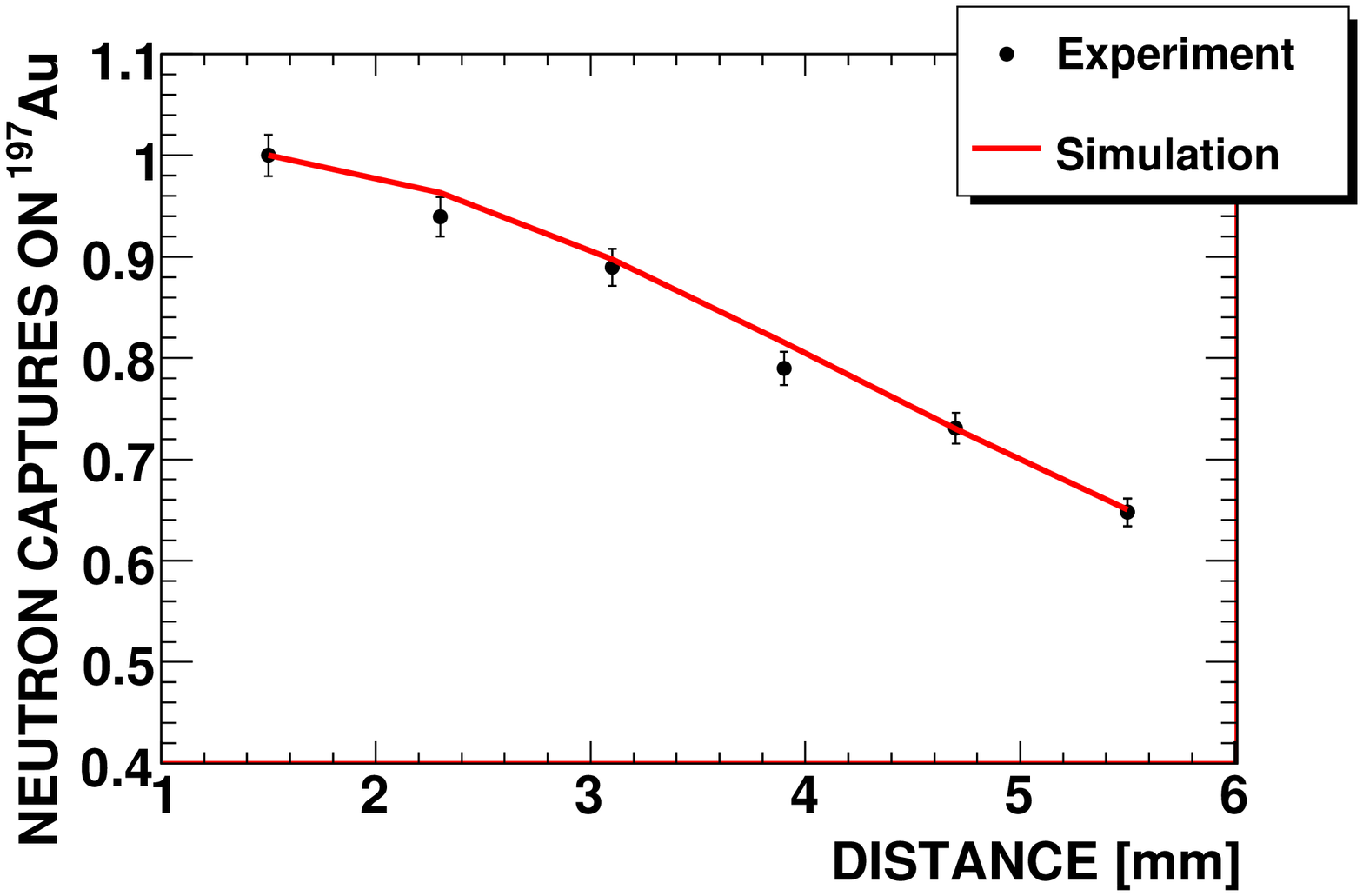}
  \caption{Left: Simulated neutron spectra for the stack of gold foils with 10~mm diameter (see also Fig.~\ref{comparison_ratynski_no_w}). Right: Comparison of
  experimental data with simulation results. The data are normalized to the point at 1.5~mm.
  \label{comparison_stack_10mm}}
\end{figure}

A second test was the comparison with an activation of a stack of gold foils \cite{RAH03}. It was found that
over a range of 6~mm, the 
induced activity in a stack of gold foils depends linearly from on the distance between the gold foils and 
the lithium spot. 
This simple behavior was found for 
gold foils of 5~and 10~mm in diameter. While the behaviour is simple, it is not so simple to explain, since the solid angle
coverage
changes non-linearly and additionally the averaged $^{197}$Au(n,$\gamma$) cross section changes because the different 
foils are exposed to different neutron spectra. It is therefore an interesting test for the simulation tool described here.
Figs.~\ref{comparison_stack_5mm} and \ref{comparison_stack_10mm} show the simulated neutron spectra and a comparison 
between experimental and simulated number of neutron captures on $^{197}$Au for the two foil diameters. The agreement 
even in these two cases is so good that we felt comfortable using this simulation tool for a number of other applications
without further verification.

\subsection{Other simulation results}

\begin{figure}
  \includegraphics[width=.9\textwidth]{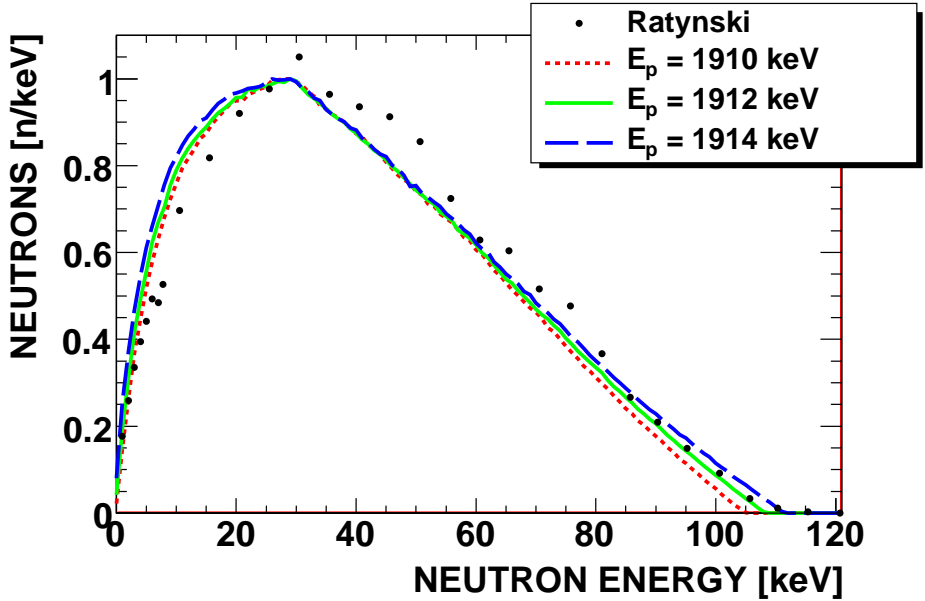}
  \caption{Comparison of the number of angle-integrated neutrons per linear energy bin for experimental results
           and simulations            
           including weighting with the angle-dependent sample thickness.
           All simulated spectra are normalized to a common maximum of 1.
  \label{comparison_ratynski_w}}
\end{figure}

In this section a few applications of the simulation tool will be presented, which have not been discussed so 
far in this paper or in previous publications. A natural first application was to simulate
the actual neutron spectrum
of the "standard" setup at the Forschungszentrum Karlsruhe. Standard setup in this case means that
the sample dimensions are such that the entire neutron cone is covered by a disk-like sample. 
Fig.~\ref{comparison_ratynski_w} shows the result of such a simulation. Obviously the simulated spectrum is
slightly shifted towards lower energies compared to the Ratynski et al. spectrum \cite{RaK88} (to be compared to
Fig.~\ref{comparison_ratynski_no_w}). The reason is 
that low-energy neutrons are preferably emitted at high emission angles and they get more weight for a 
disk-like sample.

\begin{figure}
  \includegraphics[width=.9\textwidth]{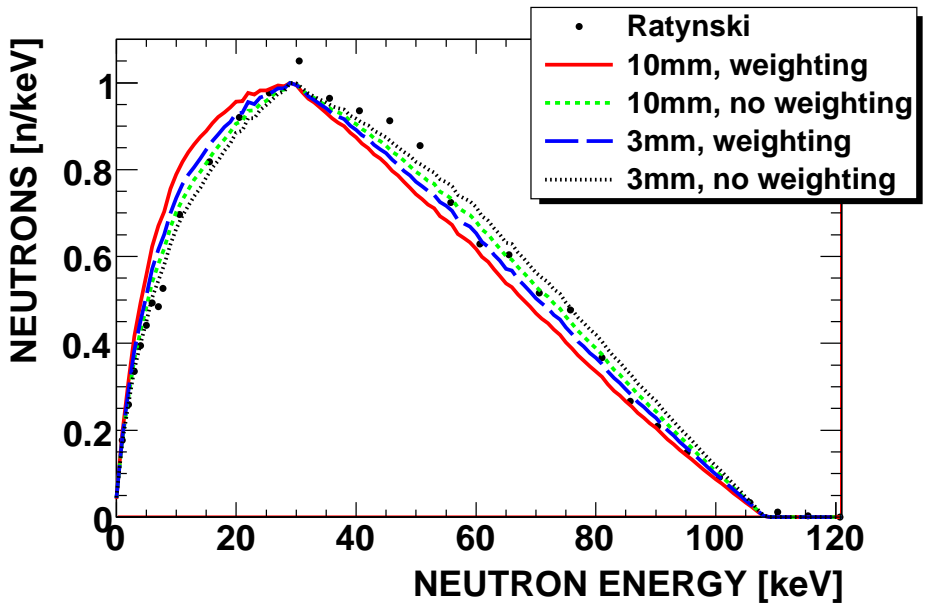}
  \caption{Comparison of the number of angle-integrated neutrons per linear energy bin for simulations that contain 
  weighting and no weighting 
  for a sample covering the entire neutron cone (10~mm radius) and only part of the cone 
  (3~mm radius). The radius of the Li-spot was 3~mm. All simulated spectra are normalized to a common maximum of 1.
  \label{comparison_ratynski_radii}}
\end{figure}

An interesting question is what happens, if only a part of the neutron cone is covered, because the sample is too small.
The question is addressed in Fig.~\ref{comparison_ratynski_radii}. The simple picture, that with smaller angular coverage
the neutron spectra are approaching the Ratynski et al. spectrum again, holds true. 

\begin{figure}
  \includegraphics[width=.9\textwidth]{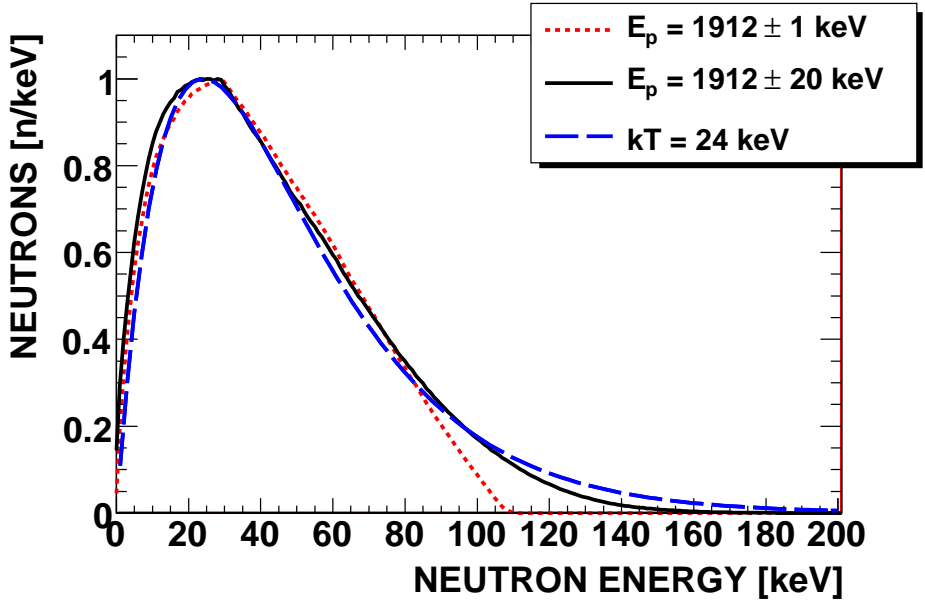}
  \caption{Comparison of the number of angle-integrated neutrons per linear energy bin for simulations 
           that contain weighting and include a gaussian proton energy profile. 
  A sample of 10~mm radius and a Li-spot of 3~mm radius was assumed. All simulated spectra are normalized to a common maximum of 1.
  \label{comparison_ratynski_gaussian_proton}}
\end{figure}

While the energy uncertainty at Van de Graaff accelerators is in the order of 0.1\%, RFQ based accelerators 
have typically uncertainties in the order of 1\%. In view of the upcoming FRANZ facility, where 
an energy uncertainty around 20~keV is expected, a simulation with the corresponding parameters has been performed.
Fig.~\ref{comparison_ratynski_gaussian_proton} shows the result of such a simulation in comparison with the much 
sharper situation at a Van de Graaff accelerator and the ideal Maxwellian averaged spectrum corresponding
to $kT~=~24~$keV. The sharp drop of the Van de Graaff spectrum at the maximum neutron energy around 110~keV is 
significantly smeared out. This means that one of the disadvantages of the activation method as applied at the 
Forschungszentrum Karlsruhe is actually almost entirely removed. A hypothetical, strong resonance at 120~keV 
would have been overlooked with the sharp edge, but would contribute at least to a certain extent at the broadened 
spectrum.    

\section{Summary} 

Driven by the needs of planning and analyzing neutron activation experiments at the Forschungszentrum
Karlsruhe, we developed a flexible tool to simulate the produced neutron spectra. 
Within the simplifying assumptions concerning proton and neutron transport the program described here is
well suited for the parameter space 
of typical activation
experiments using thin lithium layers and the $^7$Li(p,n) reaction. However, 
the code is of very restricted applicability outside of this parameter space.

We compared the resulting simulated neutron spectra with available experimental data and found
very good agreement. The program was already used undocumented during the analysis of a number 
of experiments and will also be useful for determining neutron spectra and for estimating 
neutron exposures at upcoming other facilities. 

PINO will be made available as a web-application at the URL: http://exp-astro.physik.uni-frankfurt.de/pino

\subsection*{Acknowledgments}
R.R. and R.P. are supported by the HGF Young Investigators Project VH-NG-327. We thank A.~Plompen 
for his helpful comments on the manuscript.

\newcommand{\noopsort}[1]{} \newcommand{\printfirst}[2]{#1}
  \newcommand{\singleletter}[1]{#1} \newcommand{\swithchargs}[2]{#2#1}

\end{document}